\documentclass[prd,preprintnumbers,superscriptaddress,amsmath,amssymb,nofootinbib,twocolumn,floatfix]{revtex4-2}

\usepackage{setspace}
\usepackage{graphicx}
\usepackage{hyperref}

\usepackage{units}
\usepackage{nameref, color, xcolor}
\usepackage{pifont, amsmath, amssymb, mathtools}
\usepackage{tikz}

\usepackage{booktabs}

\newcommand{\mplCircle}{\tikz\draw (0,0) circle [radius=0.9ex];}
\newcommand{\mplSquare}{\tikz\draw (-0.8ex,-0.8ex) rectangle (0.8ex,0.8ex);}
\newcommand{\mplTriUp}{\tikz\draw (0,0.95ex)--(-0.85ex,-0.75ex)--(0.85ex,-0.75ex)--cycle;}
\newcommand{\mplTriDown}{\tikz\draw (0,-0.95ex)--(-0.85ex,0.75ex)--(0.85ex,0.75ex)--cycle;}
\newcommand{\mplDiamond}{\tikz\draw (0,1.0ex)--(-0.9ex,0)--(0,-1.0ex)--(0.9ex,0)--cycle;}

\newcommand{\mplPlusFilled}{\tikz\draw[line width=0.7ex,line cap=round] (-0.9ex,0)--(0.9ex,0) (0,-0.9ex)--(0,0.9ex);}
\newcommand{\mplXFilled}{\tikz\draw[line width=0.7ex,line cap=round] (-0.85ex,-0.85ex)--(0.85ex,0.85ex) (-0.85ex,0.85ex)--(0.85ex,-0.85ex);}

\newcommand{\mplStar}{\(\star\)}
\newcommand{\mplPlus}{\(+\)}
\newcommand{\mplX}{\(\times\)}

\newcommand{\be}{\begin{equation}}
\newcommand{\ee}{\end{equation}}

\newcommand{\Basilic}{\textsc{Basilic}}

\begin{document}

\title{\textbf{Basilic: An end-to-end pipeline for Bayesian burst inference and model classification in gravitational-wave data} 
}%

\author{Iuliu Cuceu}
 \email{icuceu@oca.eu}
 \affiliation{
 Université Côte d’Azur, Observatoire de la Côte d’Azur, CNRS, Laboratoire Artemis, 06300 Nice, France}
\author{Marie Anne Bizouard}%
 \email{marieanne.bizouard@oca.eu}
  \affiliation{
 Université Côte d’Azur, Observatoire de la Côte d’Azur, CNRS, Laboratoire Artemis, 06300 Nice, France}

\date{\today}

\begin{abstract}
We present \texttt{Basilic}, a dedicated pipeline for Bayesian model selection and parameter estimation of short-duration gravitational-wave burst signals observable with ground-based detectors. Built on top of the \texttt{bilby} framework, \texttt{Basilic} combines modularity, pre-implemented burst models, and HTCondor integration to enable rapid, user-friendly analyses with minimal technical overhead. This work outlines the design philosophy, operational flow, and a set of example use cases demonstrating its scientific potential. As a case study, we also undertake an in-depth exploration of the comparison between a binary black hole merger and a cosmic string signal, through a parameter space exploration injection campaign. In addition to the well-known high-mass binary black-hole signal morphology degeneracy with cosmic string-like signals, we find that high anti-aligned component spins, even at moderate mass, can result in a similar degeneracy. Motivated by the likely low-SNR expected regime of possible future detections, we propose a data-driven study of model degeneracy, to be employed in the event of an inconclusive Bayes factor.
\end{abstract}

\maketitle


\section{\label{sec:intro}Introduction}

The LIGO-Virgo-KAGRA (LVK)~\cite{LIGOScientific:2014pky,VIRGO:2014yos,KAGRA:2020tym} collaboration has propelled the field of gravitational-wave (GW) astronomy for over a decade, with detections of compact-binary coalescences (CBCs). At the same time, a wide and astrophysically well-motivated class of short-duration transient (burst) signals remains unexplored observationally. Dedicated unmodeled searches, designed to be robust against waveform uncertainty, have been carried out across observing runs, and no confident detection of a non-CBC burst source has yet been reported in the all-sky short-duration and long-duration burst analyses of recent LVK runs~\cite{KAGRA:2021tnv,KAGRA:2025yfg,LIGOScientific:2025ksg,LIGOScientific:2014pky}. This reflects the inherently challenging nature of detecting and interpreting short-duration GW transients in nonstationary detector noise due to the difficulty of distinguishing between instrumental artifacts and potential astrophysical signals.

From an inference perspective, burst signals occupy a regime in which detection and classification decouple more sharply than for CBCs. Unmodeled (largely frequentist) burst pipelines, such as coherent WaveBurst (cWB)~\cite{drago2021coherent,Klimenko:2008fu,Klimenko:2015ypf}, X-Pipeline~\cite{Sutton:2009gi} and PySTAMPAS~\cite{Macquet:2021ttq}, are optimized to identify excess coherent power with minimal assumptions, reconstruct time-frequency structure, and deliver rapid detection statistics and sky localization. These tools are indispensable for discovery, but by construction they do not provide a complete Bayesian treatment of physical hypotheses. For short signals, this distinction matters: with few observed cycles, qualitatively different physical models can yield similar detector-frame morphologies once projected through the network response and filtered by the detector noise curve. Consequently, the relevant question for a candidate or sub-threshold event is often not only whether the data contain some coherent signal, but which of several plausible models is supported, how strongly, and with what parameter constraints.

In the burst context, Bayesian techniques~\cite{Christensen:2022bxb} are already widely used for waveform-agnostic reconstruction and glitch mitigation (e.g.\ BayesWave)~\cite{2015CQGra..32m5012C,PhysRevD.91.084034}, and past work has illustrated that physically motivated burst templates can be meaningfully classified against one another in realistic noise \cite{Divakarla:2019zjj}. For CBCs, end-to-end inference is routinely performed with standardized, production-tested pipelines built atop community libraries. For bursts, it is similarly impractical to perform multi-hypothesis Bayesian inference via bespoke scripts and ad hoc orchestration. This slows iteration, complicates reproducibility, and raises the practical barrier to systematic studies (e.g. injection campaigns), precisely in the regime where large ensembles of runs are needed to separate noise-driven ambiguity from genuine model overlap.

To address this gap, we present \Basilic~(\textbf{BA}yesian tran\textbf{SI}ent mode\textbf{L}Ing and \textbf{C}lassification), an end-to-end pipeline for Bayesian inference and model classification of short-duration GW transients, designed to complement the LVK burst-detection suite. \Basilic~is built on top of \texttt{bilby} \cite{2019ApJS..241...27A} and inherits its likelihood infrastructure and samplers, while adding the missing pipeline layer for bursts: automated run setup from a single configuration file, standardized data/PSD handling (including integration with \texttt{GWpy}~\cite{gwpy}), scalable orchestration for single-event analyses and injection campaigns, and a coherent post-processing stage that produces model-comparison summaries and diagnostic products. A central design goal is to make cross-hypothesis checks routine: a user specifies what to analyze and which models to compare, and \Basilic~handles the reproducible mechanics of running, aggregating, and reporting results. In parallel, \Basilic~ships with a growing registry of parametric burst models drawn from the literature, including core-collapse supernova phenomenological templates \cite{Powell:2022nrs,Villegas:2023wsu}, cosmic-string burst models and memory-inspired waveforms \cite{Divakarla:2019zjj}, and standard CBC approximants through \texttt{LALSuite} when burst-CBC comparisons are scientifically relevant \cite{LALSuite,SWIGLAL}.

Beyond presenting the software, this paper uses \Basilic~to produce a methodological outcome that is directly relevant for burst inference in current detectors. We perform a controlled injection campaign focused on a concrete ambiguity that has already motivated targeted analyses of heavy, short CBC-like events: the potential confusion between high-mass binary black hole (BBH) mergers and cosmic-string cusp waveforms when only a handful of cycles are in band. Such comparisons have been discussed in the context of specific events (e.g.\ GW190521~\cite{LIGOScientific:2020iuh} and GW231123~\cite{LIGOScientific:2025rsn,Cuceu:2025fzi}), and here we expand the study from single-event follow-up to a broader mapping over BBH parameter combinations at fixed detectability. In light of the results presented and burst signals in general, we also discuss explicitly some quantitative measures that can be adopted when Bayesian model selection alone is inconclusive, namely through posterior predictive checks~\cite{gelman1996posterior} and waveform match.

The remainder of this paper is structured as follows. In Sec.~II we summarize the Bayesian burst model selection problem and introduce the \textsc{Basilic} pipeline, its workflow, and the burst model catalog currently implemented. In Sec.~III we define the BBH--cosmic-string confusion study and its injection-campaign design. Sec.~IV presents the main results of the campaign, including regimes of enhanced ambiguity and their parameter dependence. Sec.~V discusses implications for short-transient interpretation and outlines how predictive-degeneracy and posterior-driven morphology studies can guide follow-up strategy. We conclude in Sec.~VI.

\section{Burst Bayesian Inference with Basilic}

Short duration GW signals are well detected by search pipelines based on coherent analysis of a multi-detector streams of strain timeseries~\cite{Klimenko:2015ypf,Sutton:2009gi,2015CQGra..32m5012C}. From the perspective of source identification, however, a Bayesian analysis performed on interesting candidates provided by such detection pipelines remains of paramount importance. We outline below Bayesian model selection and checking with \texttt{Basilic} and showcase its features that aim to bring burst model selection to an ease of use comparable to CBCs.

\subsection{Burst model selection and checking}

Unlike longer-duration signals, which can accumulate signal-to-noise ratio (SNR) through the many observed in-band cycles and thus allow for more robust model discrimination, burst source candidates are expected to have low SNR and are likely confused with noise fluctuations. Considering then the intrinsically challenging nature of a burst signal exhibiting similar signal morphology and low SNR, a complete Bayesian study is often needed.
Model selection proceeds first through Bayes' theorem, by updating the belief in  parameters $\boldsymbol{\theta}$ of a model $M$ given observed data strain $d^\ast$~\cite{Christensen:2022bxb}

\begin{equation}
    \mathcal{P}(\boldsymbol{\theta} \mid d^\ast, M) = \frac{\mathcal{L}(d^\ast \mid \boldsymbol{\theta}, M) \pi(\boldsymbol{\theta} \mid M)}{\mathcal{Z}(d^\ast \mid M)},
\end{equation}
with the a priori knowledge of said model's parameters encoded through the prior \( \pi(\boldsymbol{\theta} \mid M) \) and the Gaussian-noise GW strain likelihood \( \mathcal{L}(d^\ast \mid \boldsymbol{\theta}, M) \). To enable selection between considered models $M_1$ and $M_2$, the evidence \( \mathcal{Z}(d^\ast \mid M) \) is estimated through nested sampling~\cite{Skilling:2004pqw,10.1214/06-BA127} from which the posterior odds ratio can be computed:
\begin{equation}
    O_{12} = \frac{p(M_1 \mid d^\ast)}{p(M_2 \mid d^\ast)} = \frac{\mathcal{Z}_1(d^\ast \mid M_1)}{\mathcal{Z}_2(d^\ast \mid M_2)} \frac{\pi(M_1)}{\pi(M_2)},
\end{equation}
where in the case of equal model prior odds $\pi(M_1)=\pi(M_2)$, the odds ratio is equal to just the Bayes factor, $\mathcal{BF}_{12} = \mathcal{Z}_1(d^\ast \mid M_1)/\mathcal{Z}_2(d^\ast \mid M_2)$, which becomes the model classifier~\cite{Christensen:2022bxb}. In the case of a burst signal, models generally have differing numbers of parameters, subspaces of which can result in deceptively similar predictive waveform templates. This is despite the physically distinct nature of their sources (e.g., a linear GW memory effect versus a cosmic string cusp event).

As the Bayes factor is a measure of model ranking, the simplistic nature of many burst waveforms also motivates the use of a measure of model quality. In particular, an inconclusive Bayes factor between $M_1$ and $M_2$, does not, by itself, determine whether (i) both models adequately describe the data but are
effectively indistinguishable at the realized SNR, or (ii) one or both models are
misspecified relative to the data.
Posterior predictive checks (PPCs)~\cite{gelman1996posterior} address this by checking whether the fitted Bayesian model
\(\mathcal{M}_k(M_k,\pi_k,\mathcal{L})\) can plausibly generate data similar to the observed \(d^\ast\). For each tested model $k$, PPC requires the posterior predictive distribution for replicated datasets:

\begin{equation}
  p(d^{\mathrm{rep}}\mid d^\ast,M_k)
  =
  \int p(d^{\mathrm{rep}}\mid \boldsymbol{\theta}_k,M_k)\,
       \mathcal{P}(\boldsymbol{\theta} \mid d^\ast, M_k)\,d\boldsymbol{\theta}_k,
  \label{eq:ppc_postpred}
\end{equation}

\noindent with the replicated data $d^{\mathrm{rep}}=h_k(\boldsymbol{\theta}_k) + n^{\mathrm{rep}}$ constructed from Gaussian noise draws
colored with the representative power spectral density (PSD) of the data, and signal waveform $h_k(\boldsymbol{\theta}_k),\,\,\boldsymbol{\theta}_k \sim \mathcal{P}(\boldsymbol{\theta} \mid d^\ast, M_k)$ created from posterior parameter draws. Any tested hypothesis $M_k$ includes a nested null/noise model $M_0$, the case where \(h_0\equiv 0\). Thus, the per-model test statistic is defined by the Bayes factor against the null hypothesis:

\begin{equation}
  T_k(d)
  :=
  \log_{10}\mathcal{BF}_{k0}(d)
  =
  \log_{10}\frac{{\mathcal Z}_k(d)}{{\mathcal Z}_0(d)},
  \qquad k\in\{1,2\}.
  \label{eq:ppc_Tk}
\end{equation}

This quantifies how strongly the data support the presence of the signal family defined by model $M_k$ relative to noise alone. PPC uses the distribution of
\(T_k(d^{\mathrm{rep}})\) under \(M_k\) fitted to \(d^\ast\), and compares it to
the observed \(T_k(d^\ast)\). Thus the (one-sided) posterior predictive Bayesian p-value $p^{\mathrm{ppc}}_k$, for $M_k$, is the probability, that the test statistic of a replicated sample, $T_k(d^\text{rep})$ is larger that the observed data's, $T_k(d^\ast)$. PPC are particularly useful in the burst signal analysis scenario, as model misspecification is a real possibility. This is the case where none of the considered analysis hypotheses are actually correct with respect to the real data. Furthermore, in the case of profoundly distinct Bayesian p-values between two considered models, they can aid the model selection task by explicitly excluding the model that seems incapable of producing the observed data.

The Bayesian machinery, likelihood computation, prior selection, evidence estimate, etc., has already been implemented in GW astronomy through the python library~\texttt{bilby}. Although~\texttt{bilby} was built mainly for the most common GW detections, namely CBCs, it provides a comprehensive set of tools to allow user-defined implementations of new models, likelihoods, and other analysis utilities. These, however, require explicit scripts and extra knowledge from one looking to investigate a novel model or sub-threshold burst event. However, these require custom scripts and additional knowledge on the source studied. To address the ever-increasing number of detections and the requirement of ease of use, the CBC analysis pipeline~\texttt{bilby\_pipe} was built on top of~\texttt{bilby}, and it has been thoroughly tested in production~\cite{2020MNRAS.499.3295R}. We have built \Basilic~to provide for burst analyses, the utility that~\texttt{bilby\_pipe} provides for CBC analyses.

\subsection{Basilic}

Burst signal analyses currently face two challenges, one coming from the lack of pre-implemented and readily available burst models and one from the lack of a dedicated parameter estimation and classification pipeline for burst candidates. Thus, we have created \Basilic~an end-to-end burst signal inference and model classification pipeline, meant to complement the current LVK pipeline analysis suite. We present here its main features.

First, \Basilic~is intended to be used as an end-to-end burst inference pipeline. To facilitate this, Bayesian inference in \Basilic~is entirely handled with \texttt{bilby}, from interferometer data setup, likelihood computation and nested sampling. Furthermore, \Basilic~is HTCondor-centric~\cite{thain2005distributed}, with automatic generation of the required submission files based on the analysis type and user input, while local runs are also supported. Post-processing is managed by \Basilic~as well, with everything from run summary reports to sampling statistics, posterior parameter distributions, and model comparison figures provided. 

Second, understanding burst signals implies both offline single-event analyses and exploratory injection studies, potentially at scale. To this end, \Basilic~integrates via \texttt{GWpy} to allow up to three data input streams. Either preprocessed frame format files, or data segment downloads through \texttt{GWpy}, or simulated Gaussian noise following a chosen PSD are available. As for the PSD, it can be supplied directly through an \texttt{ASCII} format file produced by e.g., \texttt{BayesWave}, or calculated on the fly with the median-averaged power spectrum method~\cite{Allen:2005fk} as implemented in \texttt{GWpy}, or by choosing one of the many design-sensitivity PSDs pre-built into \texttt{Bilby}. A GW waveform can then be added through a simple specification of the model and parameters. If GW waveforms are added to the data, \Basilic~orchestrates both the noise realization and their seed tracking while providing summary statistics of the overall injection campaign. This campaign setup and post-processing is a central feature of \Basilic~allowing different kinds of model exploration studies, including PPC.

Third, \Basilic~provides user-friendly burst model comparison. A single configuration file defines the entry analysis for reproducibility, with only a few predefined command-line inputs required from the user after (namely submission to condor). To facilitate model comparison, \Basilic~implements multiple ready-to-use hypotheses through custom waveforms from across the burst signal literature and provides in post-processing Bayes factor tables and parameter posteriors. We expand upon the currently included burst signal models in \Basilic~in the next section.

Finally, in Appendix A, we present an example of the set of steps required to perform a single event analysis with \Basilic. We also highlight the standard outputs and post-processing that serve to facilitate the no-coding, easy to use nature of our pipeline.

\subsection{Burst models}

\begin{table*}[t]
\centering
\renewcommand{\arraystretch}{1.5}
    \begin{tabular*}{\textwidth}{@{\extracolsep{\fill}}p{0.3\linewidth} p{0.21\linewidth} p{0.27\linewidth} p{0.19\linewidth}}

    \toprule
    \toprule
    \textbf{Proto-neutron star modes} & 
    \multicolumn{3}{r}{\texttt{asymmetric\_gaussian\_chirplet}} \\
    
    \multicolumn{3}{l}{Time domain, circularly polarized}
      & \textit{Reference.:}~\cite{Powell:2022nrs} \\
    
     \multicolumn{4}{l}{
        \(
          h_+(t) = \mathcal{A}_{SN} \exp\!\left[-b\frac{(t - t_0)^2}{\tau^2}\right] \cos\left(2\pi f(t - t_0) + 2\pi \dot{f} (t - t_0)^2 + c(t - t_0)^3 \right)
        \)
      } \\
    
     \multicolumn{4}{l}{
        \(
          h_\times(t) = \mathcal{A}_{SN} \exp\!\left[-b\frac{(t - t_0)^2}{\tau^2}\right] \sin\left(2\pi f(t - t_0) + 2\pi \dot{f} (t - t_0)^2 + c(t - t_0)^3 \right)
        \)
      } \\
    
    \midrule
    
    \textbf{Parameter (Symbol)} & \textbf{Default Prior} & \textbf{Parameter (Symbol)} & \textbf{Default Prior} \\
    
    Amplitude ($\mathcal{A}_{SN}$) & $\log\mathcal{U}$($10^{-18},10^{-23}$) & Quality factor ($Q=2\pi f\tau$)  & $\mathcal{U}(10,1000)$ \\
    Peak frequency ($f$) & $\mathcal{U}$($20, 1024$) & Reference time ($t_0$)  & $\mathcal{U}$($-1,1$) \\
    Frequency derivative ($\dot{f}$) & $\mathcal{U}$($-1000,1000$) & Asymmetry ($a=\frac{1-b}{\tanh{t-t_o}}$)  & $\mathcal{U}$($-1,1$) \\
    Curvature ($c$) & $\mathcal{U}$($-500,500$) &   &  \\
    
    \bottomrule
    \toprule
    \textbf{CCSN core bounce} & 
    \multicolumn{3}{r}{\texttt{ccsn\_gaussian\_triplet}} \\
    
    \multicolumn{3}{l}{Time domain, linearly polarized }
      & \textit{Reference.:}~\cite{Villegas:2023wsu} \\
    
     \multicolumn{4}{l}{
        \(
          h_+(t) = h_1(\beta)\,
    \exp\!\left[-\frac{(t-\tau)^2}{2\eta^2}\right]
    + h_2(\beta)\,
    \exp\!\left[-\frac{(t-\tau_a)^2}{2\eta^2}\right]
    + h_3(\alpha,\beta)\,
    \exp\!\left[-\frac{(t-\tau_b)^2}{2\eta^2}\right]
        \)
      } \\

    \midrule
    
    \textbf{Parameter (Symbol)} & \textbf{Default Prior} & \textbf{Parameter (Symbol)} & \textbf{Default Prior} \\
    
    Shape parameter ($\alpha$)& $\mathcal{U}$($30,380$) & First peak arrival time ($\tau$)  & $\mathcal{U}$($-0.0005,-0.0002$) \\
    \mbox{Differential rotation} \mbox{parameter} ($\beta$)& $\mathcal{U}$($0.02, 0.14$) &  &  \\

    \bottomrule
    \toprule
    \textbf{Power law} & 
    \multicolumn{3}{r}{\texttt{generic\_power\_law}} \\
    
    \multicolumn{3}{l}{Frequency domain, linearly polarized }
      & \textit{Reference.:}~\cite{Divakarla:2019zjj}\\

     \multicolumn{4}{l}{
        \(
          \tilde{h}_+(f) = \mathcal{A}_p \Theta(f-f_{\rm low})\, \Theta(f_{\rm high}-f)\,  f^{-\alpha}
        \)
      } \\

    \midrule
    
    \textbf{Parameter (Symbol)} & \textbf{Default Prior} & \textbf{Parameter (Symbol)} & \textbf{Default Prior} \\
    
    Amplitude ($\mathcal{A}_p$)& $\log\mathcal{U}$($10^{-18},10^{-23}$) & \mbox{Low frequency} \mbox{cut-off ($f_\text{low}$)}& $\delta$($20$) \\
    Spectral index ($\alpha$)& $\mathcal{U}$($0.01,4$)& \mbox{High frequency} \mbox{cut-off ($f_\text{high}$)} & $\mathcal{U}$($25,2048$) \\

    \bottomrule
    \toprule
    \textbf{Linear memory effect} & 
    \multicolumn{3}{r}{\texttt{memory\_effect}} \\
    
    \multicolumn{3}{l}{Frequency domain, linearly polarized }
      & \textit{Reference.:}~\cite{Divakarla:2019zjj}\\

     \multicolumn{4}{l}{
        \(
          \tilde{h}_+(f) = -i\pi \mathcal{A}_m \Theta(f-f_{\rm low})\, \Theta(f_{\rm high}-f)\,\tau  / \sinh{(\pi^2\tau f)}
        \)
      } \\

    \midrule
    
    \textbf{Parameter (Symbol)} & \textbf{Default Prior} & \textbf{Parameter (Symbol)} & \textbf{Default Prior} \\
    Amplitude ($\mathcal{A}_m$)& $\log\mathcal{U}$($10^{-18},10^{-23}$) & \mbox{Low frequency} \mbox{cut-off ($f_\text{low}$)}& $\delta$($20$) \\
    Damping time ($\tau$)& $\log\mathcal{U}$($10^{-4},10^{-1}$)& \mbox{High frequency} \mbox{cut-off ($f_\text{high}$)} & $\mathcal{U}$($25,2048$) \\
    
    \bottomrule
    \toprule
    \textbf{Damped sinusoidal transient} & 
    \multicolumn{3}{r}{\texttt{damped\_sine}} \\
    
    \multicolumn{3}{l}{Time domain, linearly polarized}
      & \\

     \multicolumn{4}{l}{
        \(
        \tilde{h}_+(f) = \frac{A_g\tau}{2}\exp\!\left(-2\pi i f t_0\right)\big[(1+2\pi i\tau(f-f_0))^{-1}+(1+2\pi i\tau(f+f_0))^{-1}\big]
        \)
      } \\

    \midrule
    
    \textbf{Parameter (Symbol)} & \textbf{Default Prior} & \textbf{Parameter (Symbol)} & \textbf{Default Prior} \\
    Amplitude ($\mathcal{A}_g$)& $\log\mathcal{U}$($10^{-18},10^{-23}$) & Initial frequency ($f_0$)& $\mathcal{U}(25,1024)$ \\
    Damping time ($\tau$)& $\log\mathcal{U}$($10^{-4},10^{-1}$)&  Reference time ($t_0)$ & $\mathcal{U}(-1,1)$ \\
    \bottomrule
    \bottomrule 
    \end{tabular*}
  
\caption{All the currently implemented models in \Basilic~beyond those available in \textsc{bilby}. The name and \Basilic~designation (right) are specified. For each model, we mention both the waveform template and the default priors on the free parameters. $\mathcal{U}$ denotes a uniform distribution, $\log\mathcal{U}$ a log-uniform, $\delta$ a delta function or a fixed parameter.}
  \label{tab:models}
\end{table*}

One of the main challenges with burst signals comes from the potential similarity of signal morphology. To enable Bayesian analysis and model classification of said signals, a catalog of waveforms needs to be readily available. From a user perspective, this means a one-line change in the configuration file choosing what model to test. Moreover, \Basilic~utilizes a modular approach, which allows for easy future implementation of any parametric gravitational waveform to be added to the existing catalog.

At the time of writing, \Basilic~includes the entire list of sources available in \textsc{bilby}, as well as some added models to round up a burst signal search. On the CBC side, all waveforms currently implemented in \textsc{lalsuite} are also available in \Basilic~through a change of the waveform approximant argument in the configuration. Furthermore, through \textsc{bilby}, we include functionality for simple Sine-Gaussian and Damped-Sinusoid models. 
Each model has a set of default priors that can be customized.
To round up our catalog of source models, \Basilic~includes support for two core collapse supernova (CCSN) models, three cosmic string models, a memory effect, a generic power law, and a half-damped sine glitch model. On the supernova front, we have a minimally modeled core-bounce phase that appears in rapidly rotating CCSNe, the three-peak template proposed in~\cite{Villegas:2023wsu}. This model is meant to capture the short ($\lesssim10\,\mathrm{ms}$) axisymmetric bounce signal with the smallest number of parameters needed for inference on rotation and equation of state (EOS) proxies. For the period after the core bounce, we implement a phenomenological model introduced in~\cite{Powell:2022nrs}. It tries to capture the dominant emission components, the high-frequency gravitational ($g$)- and fundamental ($f$)-modes associated with oscillations of the proto-neutron star surface, as well as low-frequency standing accretion shock instability mode. 

The three main single-event cosmic string sources are cusp, kink and kink-kink string interactions, which can all be modeled in the frequency domain by a power law, with different spectral indices: $-4/3$ for cusp, $-5/3$ for kinks and $-2$ for kink-kink. We include these three, as well as a more generic power law model (varying spectral index), to capture a range of possible signals. A power law is also the limiting case (rise time goes to zero) of the linear gravitational memory effect, a slow-rise, broadband, and non-oscillatory signal associated with the time-varying spacetime change when GWs are generated. The cosmic string, power law, and memory effect burst models implemented here are introduced in~\cite{Divakarla:2019zjj}. Finally, motivated by the brief-in-time, peaked-in-frequency nature of many instrument artifacts, we also include a time-domain half-damped sine model, which we implement directly in the frequency domain to avoid Fourier transform frequency leakage. A summary list of the newly added models that have been taken from the burst model literature is available in Table~\ref{tab:models}.

\section{Case study, BBH vs cosmic string}

\begin{table*}[t]
\centering
\renewcommand{\arraystretch}{1.2}
\begin{tabular}{@{}lccccccccc@{}}
\toprule
\textbf{Label} && $\boldsymbol{\mathcal{M}_c}~[\mathrm{M}_\odot]$ & $\boldsymbol{q}$ & $\boldsymbol{a_1}$ & $\boldsymbol{a_2}$ & $\boldsymbol{\theta_1}$ [deg] & $\boldsymbol{\theta_2}$ [deg] & $\boldsymbol{d_L}$ [Mpc]& \textbf{Plot marker}\\
\midrule
A1  && 71.536 & 0.85 & 0.6 & 0.6 & $15\pi/16$ & $15\pi/16$ & 6500 & \mplCircle \quad\\
A2    && 71.536 & 0.85 & 0.2 & 0.2 & $\pi/16$ & $\pi/16$ & 8000& \mplSquare  \quad \\
\midrule\midrule
B1 && 86.712 & 0.85 & 0.9 & 0.9 & $15\pi/16$ & $15\pi/16$ & 9200&\mplTriUp  \quad\\
B2  & & 86.712 & 0.85 & 0.2 & 0.2 & $\pi/16$ & $\pi/16$ & 6900&\mplTriDown  \quad \\
B3& & 74.341 & 0.35 & 0.9 & 0.9 & $15\pi/16$ & $15\pi/16$ & 7600& \mplDiamond  \quad \\
B4  &    & 74.341 & 0.35 & 0.2 & 0.2 & $\pi/16$ & $\pi/16$ & 6600& \mplPlusFilled  \quad\\
\midrule\midrule
C1 & & 108.344 & 0.85 & 0.6 & 0.6 & $\pi/2$ & $\pi/2$ & 8900& \mplXFilled  \quad\\
C2 & & 108.344 & 0.35 & 0.6 & 0.6 & $\pi/2$ & $\pi/2$ & 6500& \mplStar  \quad\\
\midrule\midrule
D1 & & 173.424 & 0.85 & 0.9 & 0.9 & $\pi/2$ & $\pi/2$ & 11600& \mplPlus  \quad\\
D2  &    & 173.424 & 0.85 & 0.2 & 0.2 & $\pi/16$ & $\pi/16$ & 13000& \mplX \quad\\
\midrule
\textbf{Prior} & &$\mathcal{U}^\star[70,200]$ & $\mathcal{U}^\star[1/6,1]$ & $\mathcal{U}[0,0.99]$ & $\mathcal{U}[0,0.99]$ & $\mathrm{Sine}(0,\pi)$ & $\mathrm{Sine}(0,\pi)$ & $\mathcal{U}^\star[500,1500]$\\
\bottomrule
\end{tabular}
\caption{List of parameters (chirp mass, mass ratio, component spin amplitudes, component spin alignment, and luminosity distance, respectively) for the BBH systems used to showcase the apparent confusion between high-mass BBH and cosmic string signals. The row segmentation is made by total mass.  The last row lists the priors used to sample the parameters.
The star asterisk for chirp mass, mass ratio and luminosity distance indicates that in fact \texttt{bilby} implements a uniform-in-components chirp mass, uniform-in-components mass ratio and uniform-in-source frame priors. For a description of these priors, see~\cite{2019ApJS..241...27A}. }
\label{tab:injections-compact}
\end{table*}

To showcase \Basilic~in action and as the main scientific outcome, we present in this section the setup of a BBH and cosmic string model comparison study, and in the next section, the main results of this work. Utilizing the signal injection facility of \Basilic, we investigate how certain BBH parameter combinations can result, through a Bayesian model comparison lens, in cosmic string-like signals. 

Burst signal morphology similarity (especially in the presence of noise) is partially a consequence of the signal duration. Since no current above-threshold GW event is likely attributed to a burst source~\cite{KAGRA:2021tnv,KAGRA:2025yfg,LIGOScientific:2021nrg}, there has naturally developed an interest in how known short-duration GW sources (namely BBH) could appear as burst signals instead.

A particular case of this is the apparent high-mass BBH-cosmic string potential source of confusion. As the total mass of a binary system increases, the maximum frequency (merger frequency), and thus the maximum detectable GW frequency, decreases. This effect, coupled with the characteristic chirp-like frequency evolution of CBC signals and the rise in detector noise at low frequency ($\sim 20$ Hz) results in shorter observable signals. Therefore, when a particular GW event is best fit by a high-mass BBH waveform template, it is natural to check how said event is fit by a cosmic string waveform template as well. This is in fact, the motivation behind performing Bayesian model comparison studies between BBH and cosmic string in the highest mass BBH GW candidates, as was done with GW190521~\cite{LIGOScientific:2020iuh} and GW231123~\cite{Cuceu:2025fzi}. We thus aim to expand~\cite{Cuceu:2025fzi}, investigating this known part of the BBH parameter space that results in ambiguity with cosmic string signals by studying other parameter combinations.
Table~\ref{tab:injections-compact} summarizes the intrinsic parameters of BBH considered in this study, grouped by total mass. We focus our study on the effect of varying the chirp mass, mass ratio, spin amplitudes, and spin alignment parameters. Finally, we note that we have also performed a smaller-scale mirror study (cosmic string injections, with BBH recovery) that we do not include here. This is because we believe that the larger interest is in BBH parameter properties that can lead to confusion, as BBH is already an observationally confirmed GW source. Furthermore, since short BBH signals can lead to confusion with multiple burst models, this case study should be used to highlight the limitations and strengths of BBH Bayesian parameter estimation, rather than the intrinsic difficulties with Bayesian methods or burst models at low SNR. This former point is discussed in detail in Sec.~\ref{sec:discussion}.

We consider a network formed by the two LIGO detectors, Hanford (H1) and Livingston (L1) for which Gaussian noise is generated using the LIGO design sensitivity. For each BBH parameter choice, we inject the signal waveform in one hundred simulated Gaussian noise realizations.

To facilitate a fair comparison, we keep the optimal SNR of the network constant ($\rho_\text{opt}=6$) when choosing the luminosity distance of the source. Such a low SNR corresponds to a regime where reconstructed transient signals are affected by noise. We consider the BBH waveform approximant \texttt{NRSur7dq4}~\cite{Varma:2019csw}, because of its accuracy in the highly spinning, high-mass CBC signals.

For each BBH signal plus noise realization, we carry out a full Bayesian model comparison under two hypotheses: the true injected one, \texttt{NRSur7dq4} approximant, and a cosmic string cusp (CSC) waveform. In practice, each hand-picked BBH parameter combination is simply a different \texttt{YAML} configuration file provided to \Basilic, which alternatively can also be provided with parameter grids in a single setup file as well, the injection campaign is then automated and results aggregated. For reproducibility, we utilize \unit[4]{s} segments, a sampling frequency of \unit[2048]{Hz} and $600$ live points with the nested sampler \texttt{dynesty}~\cite{2020MNRAS.493.3132S,koposov:2023}.

Since our study focuses on combinations of BBH parameters that resemble cosmic string cusps in general, our CSC prior is quite unconstrained, with the amplitude log-uniform in $[10^{-23},\,10^{-18}]$ and high frequency cutoff uniform in $[25, \,512]$. As for the sky localization, we assume the position of the BBH source in the sky to be known, to lower problem dimensionality, while we let the cosmic string search over the whole sky. The temporal localization is uniform $\pm1$\,s around the injection for all cases. We report that under tests with different spatio-temporal prior choices of the considered hypotheses, the resulting Bayes factor distributions were not affected. The luminosity distance is chosen per injection to yield network optimal SNR of 6 in the H1-L1 network configuration. Furthermore, we fix all the other BBH parameters not mentioned in Table~\ref{tab:injections-compact} to their injected values with delta function priors, namely $\theta_\text{jn}=\pi/4,\phi=\pi/2,\Delta\phi=\pi,\phi_\text{JL}=\pi/2,\alpha=1,\delta=2,\psi=0$. For a description of these priors, see~\cite{2019ApJS..241...27A}. Fixing these parameters was done to keep the computational cost of the $10 \times 100$ estimations of the BBH evidence low. In contrast, the lower dimensionality of the CSC model made their evidence estimation negligible in comparison. We further wish to mention that restrictive priors do affect the evidence estimation and subsequent Bayes factor recovery. However, less restrictive priors can only lower the support for the BBH hypothesis, thus strengthening the confusion effect. Therefore, due to computational constraints, we have excluded this from our study.

\section{Results}

\begin{figure}
    \centering
    \includegraphics[width=1\linewidth]{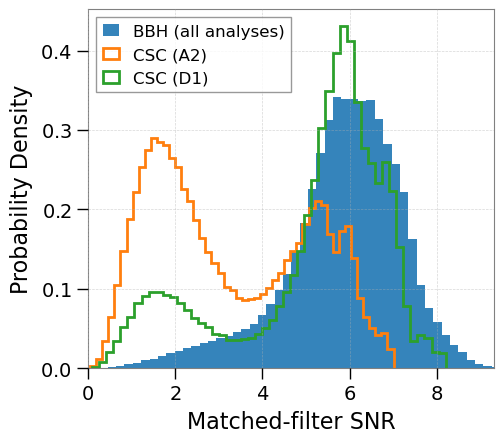}
    \caption{SNR density, constructed from aggregating all the per-posterior sample matched-filter SNRs, under the BBH hypothesis, of all noise realizations and of all signal injections are overlaid in blue. In orange and green are the SNR distributions under the CSC hypothesis of the A2 and D1 analyses, respectively. These two are also the aggregated SNRs over all noise realizations of the analyses in question. }
    \label{fig:SNRs}
\end{figure}

Figure \ref{fig:SNRs} displays posterior distributions of matched filter SNR across all noise realizations under the BBH (true) hypothesis. In more detail, the blue histogram shows the aggregated SNR posteriors for all BBH injections. The distribution is centered $\sim6$, since the luminosity distance is chosen to result in an optimal SNR of 6 for each injection. The lower-end SNR part is predominantly due to noise effects and, to a lesser extent, to the quality of the nested sampling parameter estimation.

\begin{figure*}[t]
  \centering

  \begin{minipage}{0.49\textwidth}
    \centering
    \includegraphics[width=\linewidth]{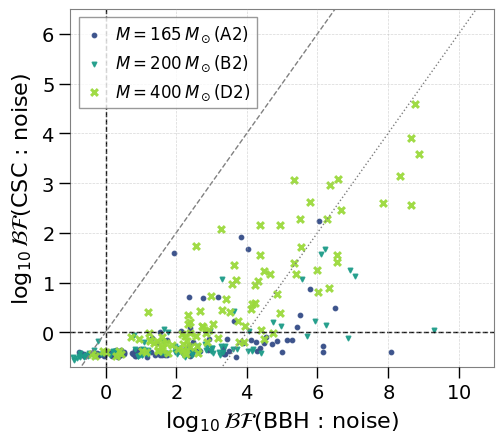}
  \end{minipage}\hfill
  \begin{minipage}{0.49\textwidth}
    \centering
    \includegraphics[width=\linewidth]{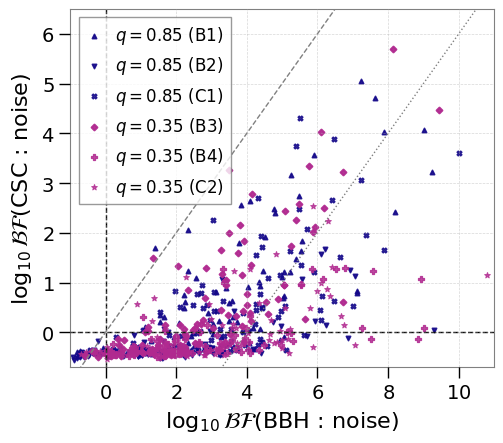}
  \end{minipage}

  \vspace{0.75em}

  \begin{minipage}{0.49\textwidth}
    \centering
    \includegraphics[width=\linewidth]{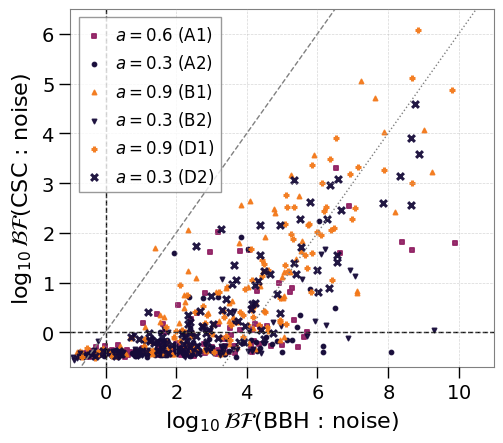}
  \end{minipage}\hfill
  \begin{minipage}{0.49\textwidth}
    \centering
    \includegraphics[width=\linewidth]{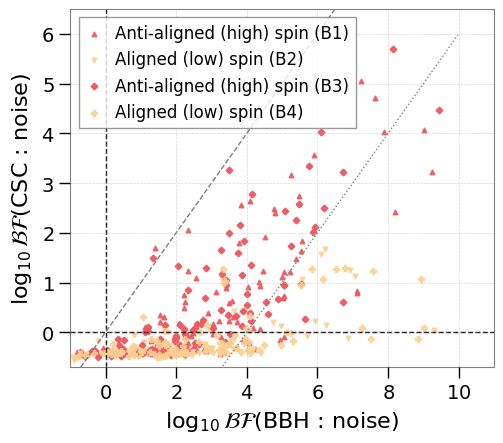}
  \end{minipage}

  \caption{Bayes factor vs Bayes factor plots of various analysis combinations, comparing the evidence for the BBH hypothesis (x-axis) and cosmic string cusp hypothesis (y-axis) relative to the no-signal hypothesis. Each analysis has a distinct marker style and shape (see Table~\ref{tab:injections-compact}), while we vary the colors of these markers based in each plot. Top left, we highlight the effect of BBH mass variation with the considered analyses (A2 in blue, B2 in light blue and D2 in lime) having only their total mass differing. Top right, B1, B2 and C1 analyses in navy blue represent the high mass ratio regime, while B3, B4 and C2 in violet, the low mass ratio regime. Bottom left, we color by spin magnitude, with A2, B2 and D2 analyses ($a=0.3$) in dark blue, A1 ($a=0.6$) in purple and B1 and D1 ($a=0.9$) in orange. Bottom right, the effect of spin alignment is highlighted, with (low) aligned spin analyses B2 and B4 in light orange, and the (high) anti-aligned spin B1 and B3 in light red.}
  \label{fig:BF_grid}
\end{figure*}

We note that, had the injections been closer (higher SNR), the width of the recovered overall distribution would decrease. In other words, noise variation plays a smaller role as the signal strength increases. The variance of the SNR posterior distribution under the true model hypothesis is important in showcasing the difficulties in studying the low-SNR regime. In particular, as it pertains to model selection, we contrast the recovered BBH SNR distribution with two example distributions under the CSC hypothesis represented on the same figure in orange and green for low (A2 analysis) and high-mass (D1 analysis) BBH injections respectively.
We see clearly, that in the known high-mass BBH vs cosmic string confusion scenario, the recovered SNR distributions are closer, while for the low mass BBH injection, testing the CSC hypothesis results in a consistently low SNR. The distinct, two-peak nature of these distributions is due to noise characteristics of a given injection versus another. In particular, when the injected BBH signal (D1 analysis) does resemble a cosmic string, most noise realizations result in recovered CSC SNR distributions similar to those of BBH (one peak), with some outliers that smear out the distribution to lower values (where parameter estimation quality suffers as well). In contrast, the reverse is true for a less model-ambiguous injection like A2, where most samples result in CSC SNR distribution clustering around lower values, with outliers on the higher end.

Moving on from the SNR, the Bayes factor is the usual Bayesian model classifier. We thus study the distributions of Bayes factors across noise realizations for each BBH signal, in an attempt to shed some light on the possible regimes of confusion between our considered models. For every injection and noise realization, we analyze the Bayes factor of a tested signal hypothesis (BBH or CSC) versus the noise hypothesis (no signal present).
Figure~\ref{fig:BF_grid} shows scatter plots of $\log_{10}({\mathcal BF}_{\rm BBH})$ and $\log_{10}({\mathcal BF}_{\rm CSC})$, where each point is from a given noise realization and injection; the marker shape identifies the injection parameters, while we vary the colors for plot interpretability. As a general note on such plots, a $\log_{10}({\mathcal BF})>8$ is considered strong evidence in GW astronomy and values below zero favor the no-signal hypothesis, which we delineate with dashed black lines. Furthermore, direct model comparison (between BBH and CSC) would represent a projection on some $x-y=s$ axis with conventional choices for $s$ as the decision boundaries for model selection~\cite{kass1995bayes}. Lastly, we vary the colors based on the parameter considered for each plot, with the marker symbol constant throughout for each analysis and found in Table~\ref{tab:injections-compact}.

First, we highlight the effect of varying the total mass (all other parameters are equivalent) of a BBH injection and the resulting problem of model indistinguishability in the top left of Figure~\ref{fig:BF_grid}. All three injections, A2, B2 and D2, in blue, light blue and lime respectively, display similar model distinguishability (moderate evidence) for BBH hypothesis relative to no-signal hypothesis. However, the high-mass D2 analysis displays a clearer correlation between $\log_{10}({\mathcal BF}_{\rm BBH})$ and $\log_{10}({\mathcal BF}_{\rm CSC})$. Therefore, for these BBH injections, depending on the noise realization, one can neither claim ``not a CSC signal'' nor ``not a BBH signal''. In contrast, for a lower-mass BBH injection, most cases ($~90\%$ for A2) prefer the no-signal hypothesis to the CSC signal one. 

Next, we investigate the impact of the mass ratio. The top right of Figure~\ref{fig:BF_grid} shows six pairwise-equivalent analyses (B1-B3, B2-B4, C1-C2), with variation only in the component mass ratio, in navy blue ($q=0.85$) and violet ($q=0.35$). No separation is observed by color and thus by mass ratio, unlike the previous case by total mass. In other words, for the cases considered here, B1 through C2, similar or different component mass combinations, with all other parameters constant, does not drive model confusion, or lack thereof with cosmic string. 

In the bottom left of Figure~\ref{fig:BF_grid} we vary the spin magnitude, $a=0.3$, $0.6$ or $0.9$, in dark blue, purple, and orange, respectively. The six pairwise analyses (A1-A2, B1-B2, D1-D2) tell a similar story; spin magnitude alone does not seem to drive model confusion between BBH and cosmic string. We note, however, that by design, we considered anti-aligned spin injections to also be high spin magnitude ones. Therefore, spin amplitude alone does not serve as complete a role of explanatory variable, as opposed to the mass ratio. Grouping the analyses by variation in spin alignment, on the bottom right of Figure~\ref{fig:BF_grid}, with aligned spins in light orange and anti-aligned in light red, does exhibit a clear separation again. Similar to the case of varying the total mass, we see that anti-aligned (high) spins result in a stronger correlation between the Bayes factors, thus making the injections "look more like a cosmic string". 

Finally, we note the presence of dense clusters in the $\log_{10}(\mathcal{BF}_\text{CSC})<0$ region for all considered injections; although these cases do correspond to ``not a CSC signal'', injections with $\log_{10}(\mathcal{BF}_\text{BBH})\lesssim3$ at best permit a ``moderate evidence for BBH'' interpretation. In other words, this is precisely the regime where ``unlucky'' noise realizations do not permit model selection anyway, due to low SNR.
To go further, we should contextualize this dense region, with the comparatively empty region where $\log_{10}(\mathcal{BF}_\text{CSC})<0$ and $\log_{10}(\mathcal{BF}_\text{BBH})>5$. The impact of individual parameters on model confusion, is not best showcased by the value of the strength of the correlation of any given parameter set in isolation, since loud noise buries any signal, so the shape of the distribution is irrelevant for model confusion in a high-noise-evidence regime. Rather, it is instructive to look at the change of the coefficient of correlation $r$ of the Bayes factors when varying parameters of interest. For example, A2 corresponds to $r_\text{A2}=0.39$, while D2 has $r_\text{D2}=0.87$, showing the change when total mass grows. Similarly, for the spin alignment case, aligned low spin B4 has $r_\text{B4}=0.61$ while anti-aligned high spin B3 corresponds to $r_\text{B3}=0.78$. In contrast, varying the mass ratio, low mass ratio B3 versus high mass ratio B1 results in $r_\text{B3}-r_\text{B1}=0.04$. These results, highlight the fact that anti-aligned high component spins, as well as the known, high component masses of BBH, both result in burst-like signals that could lead to confusion in model selection.

\section{Discussion}\label{sec:discussion}
The work presented here, on certain BBH parameter combinations that can lead to confusion with burst models (here cosmic string models), is the result of large-scale injection campaigns run with \Basilic. It is important to note, that in the absence (or very low level) of noise, neither a high-mass BBH nor a highly spinning BBH, as presented above, result in waveforms that are similar enough to ``fool'' a Bayesian analysis. However, in the presence of noise (here, Gaussian following the advanced LIGO sensitivity), it seems clear, that there exist parameter combinations for BBH waveforms (\texttt{NRSur7dq4} approximant), that are particularly prone to being underclassified (via Bayes factor) in relation to CSC waveforms. 

We have observed instances (e.g. D2, B1 analyses) where the noise Bayes factors of cosmic string signals grow as the noise Bayes factor of the BBH hypothesis. For other BBH parameters (e.g A1, B2) the cosmic string Bayes factor stays consistently in favor of noise, irrelevant of the BBH Bayes factor. This indicates the existence of BBH parameter space regions that are ``degenerate'' with a CS cusp waveform. In other words, given multiple (lower SNR) observations that seem to correspond to BBH within that parameter space, the Bayes factor comparing the two signals hypothesis, $\log_{10}\mathcal{BF}_\text{BBH-CSC}$ stays consistently low, relative to some decision boundary. Alternatively, from an SNR perspective, we can say that certain parameter regions require higher SNRs to provide the discrimination power necessary for model selection.

These results highlight a clear consequence of analyconfusionzing short-duration signals. With bursts, there are only so many ways to create sufficiently different waveforms that result in clear discriminative power in their subsequent Bayes factor at moderate SNR. Observing such degeneracy effects even in the case of BBH and CS, where there is a priori sufficient discriminative power (through e.g. BBH possessing a clear chirping signal in general, vs CS that do not), showcases the need for careful model checking for all burst signal hypotheses. 

Since this low-SNR, high potential for model degeneracy regime seems probable for burst-like data and burst signal combinations, we propose a strategy, both to aid model selection and to quantify degeneracy. Starting with an observation $d^\ast$ and two models we consider viable for testing $M_1,M_2$:
\begin{enumerate}
\item \textbf{Bayesian model comparison:}
Compute evidences for noise $Z_0(d^\ast)$, signals $Z_1(d^\ast)$, $Z_2(d^\ast)$ and Bayes factors ${\mathcal BF}_{12}$, ${\mathcal BF}_{10}$, ${\mathcal BF}_{20}$. An inconclusive
\(\mathcal {BF}_{12}\) together with strong/decisive \(\mathrm{BF}_{10}\) and \(\mathrm{BF}_{20}\)
indicates a confident signal detection but ambiguous classification.

\item \textbf{Stage I PCC (model adequacy):}
For each signal hypothesis \(k\in\{1,2\}\), draw replicated datasets
\(d^{\mathrm{rep}}\sim p(d^{\mathrm{rep}}\mid d^\ast,\mathcal{M}_k)\) and compute \(p^{\mathrm{ppc}}_k\) from Eq.~\eqref{eq:ppc_postpred}.
If one model yields an extreme Bayesian p-value while the other does not, this supports the interpretation that the latter model is less capable of generating data like \(d^\ast\).

\item \textbf{Stage II match-based degeneracy (morphology):}
If PPC still does not clearly separate the models, quantify posterior-conditioned waveform
degeneracy via an analysis of the waveform match. A complete description of this match analysis can be found in Appendix~\ref{app:pdeg}.
High levels of waveform match across parameter combinations, in the posterior-relevant
regions, indicate that the waveform families are morphologically
degenerate; low values indicate that the inconclusive \(\mathrm{BF}_{12}\) is more likely
driven by finite-SNR/noise fluctuations than by intrinsic waveform similarity.
\end{enumerate}

With the help in orchestration by \Basilic~these further checks are trivial. This exact scheme is currently implemented in the postprocessing module as optional checks in the inconclusive model selection case.

\section{Conclusion}

We have presented \Basilic~(BAyesian tranSIent modeLIng and Classification), an end-to-end pipeline for Bayesian inference and model comparison of short-duration gravitational-wave transients. Built on top of \texttt{bilby}, \textsc{Basilic} provides the burst pipeline layer that makes cross-hypothesis analyses routine: a single configuration file defines data and PSD handling, signal hypotheses and priors, and the execution mode (single-event or injection campaign), while HTCondor orchestration and standardized post-processing produce reproducible results and curated diagnostic outputs~\cite{Cuceu2025}.

We used \Basilic~to address a concrete ambiguity that has motivated targeted analyses of massive, short CBC-like events: potential confusion between high-mass BBH signals and cosmic-string cusp templates when only a few cycles are in band. Using a controlled injection campaign at fixed SNR, over a set of representative BBH configurations and many independent noise realizations, we mapped how evidence-level behavior changes as intrinsic BBH parameters vary. The results show that the tendency for the cosmic string evidence to track the BBH evidence strengthens markedly with increasing total mass, and also increases for anti-aligned (high) spin configurations, while the mass-ratio variations explored here do not produce an analogous separation. Importantly, the observed ambiguity is noise-enabled: in the absence of (or at very low levels of) noise, the BBH and cosmic string waveform families naturally remain sufficiently distinct, whereas noise fluctuations can render specific BBH regions prone to underclassification against cosmic string waveforms.

These findings sharpen how we should interpret Bayes factors for short transients in practice. First, in the low-SNR regime, many noise realizations yield at most moderate support for any signal model, so an absence of strong model preference can be driven primarily by detectability rather than intrinsic degeneracy. Second, when the signal is sufficiently supported over noise yet the Bayes factor between two physical hypotheses remains inconclusive, the ambiguity can arise either from genuine morphological overlap of the posterior-supported waveform families or from noise fluctuations. To operationalize this distinction, we outlined (and implement in \Basilic~post-processing) a two-stage diagnostic strategy for inconclusive comparisons: PPCs to test model adequacy, followed, when needed, by a posterior-conditioned, match-based degeneracy criterion that quantifies whether high-match counterparts exist across the competing waveform families.

\Basilic~is intended to support both single-event follow-up and systematic injection-based exploration of burst inference in the LVK. The pipeline development will continue beyond the features presented here, with the current goals being: the expansion of the model registry and post-processing diagnostics. Currently, \Basilic~provides both a practical analysis framework and a scalable route to quantifying where burst model selection is robust, where it is noise-limited, and where waveform-family construction limitations genuinely constrain astrophysical interpretation.

\begin{acknowledgments}
We would like to thank Renate Meyer and Nelson Christensen for their significant insights.
The authors are grateful for computational resources provided by the LIGO Laboratory and supported by the National Science Foundation Grants PHY-0757058 and PHY-0823459. This material is based upon work supported by NSF’s LIGO Laboratory which is a major facility fully funded by the National Science Foundation. This research has made use of data~\cite{KAGRA:2023pio} or software obtained from the Gravitational Wave Open Science Center (gwosc.org), a service of the LIGO Scientific Collaboration, the Virgo Collaboration, and KAGRA.

\end{acknowledgments}

\appendix

\section{Single event example: GW231123}\label{app:gw231123}
We demonstrate the steps required to run \Basilic~and its single-run output, on a real data event analysis, GW231123~\cite{LIGOScientific:2025rsn}. As mentioned in the introduction, the analysis in~\cite{Cuceu:2025fzi} was performed with \Basilic. The work in~\cite{Cuceu:2025fzi} investigates the origin of the GW231123 event, comparing  binary black hole, three cosmic string models, and a generic power law model. Here we report only the information relevant to this paper, namely, the setup and output of \Basilic.

After installation of the \Basilic~pipeline\footnote{Available on GitLab at~\cite{Cuceu2025}, along with installation instructions, requires LVK credentials, for the moment.}, the analysis requires an initial configuration file (\texttt{YAML} format), setting up the data, the PSDs and the hypotheses to test. For GW231123,~\cite{Cuceu:2025fzi} utilizes a PSD provided by  \texttt{BayesWave}~\cite{Cornish:2014kda}, in \texttt{ASCII} format, and analysis-ready interferometer strains downloaded from the GWOSC website~\cite{gwoscGW231123}, so the configuration file simply points to the direction of the data and PSD sources. The mandatory arguments in the configuration files are \texttt{detectors}, \texttt{trigger\_time}, \texttt{data\_source}, \texttt{psd\_source}, and \texttt{model\_names}, along with some Condor job submission. A full list of possible arguments, their defaults, requirement status and instructions, along with the complete configuration file for the GW231123 analysis in~\cite{Cuceu:2025fzi} are available at~\cite{Cuceu2025}.

\begin{figure*}
    \centering
    \includegraphics[width=1\linewidth]{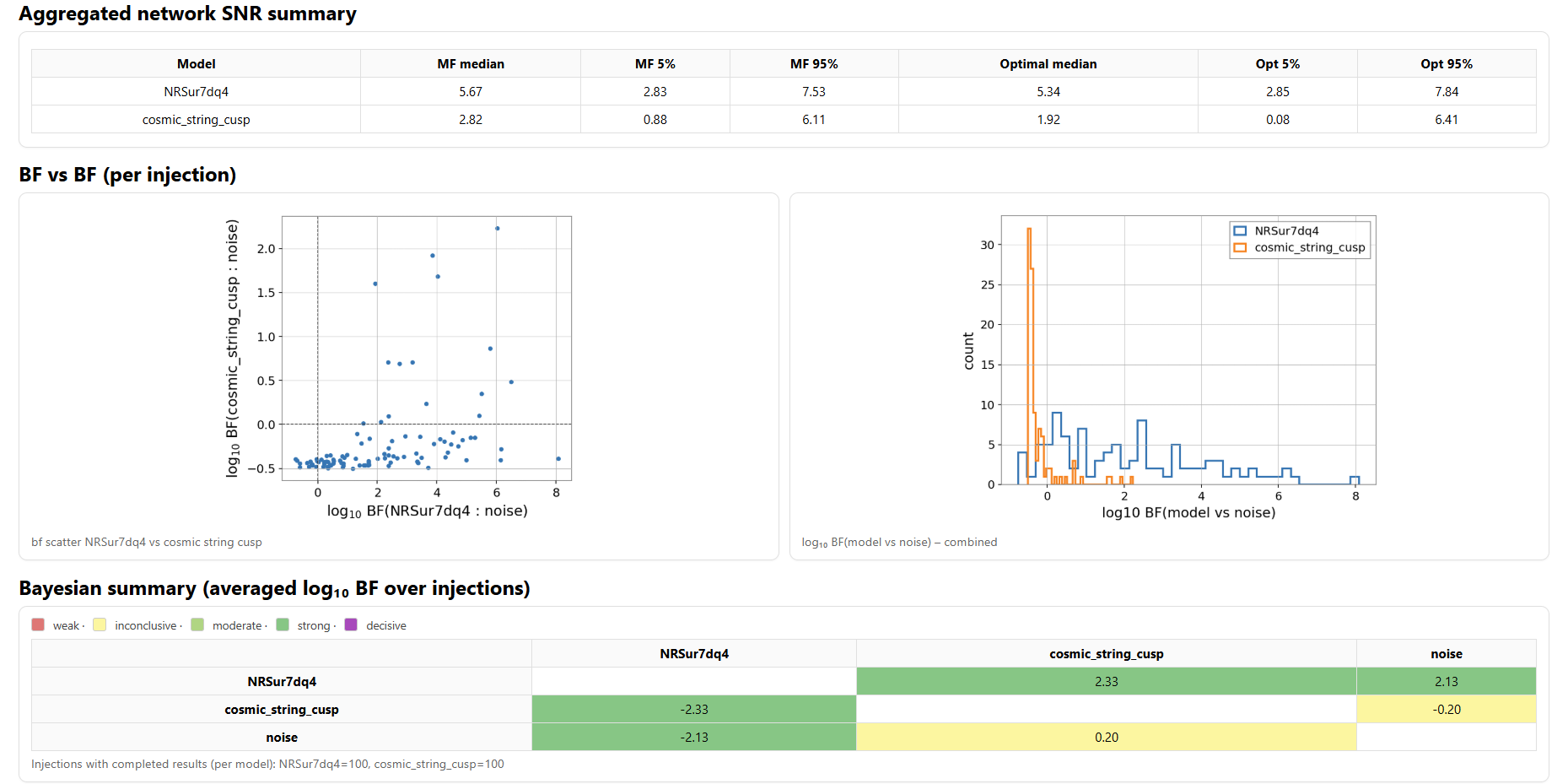}
    \caption{Part of the \textsc{HTML} report generated output by \Basilic. The format and plots are different in the injection campaign versus single-event runs.}
    \label{fig:screenshot}
\end{figure*}

Having a written configuration file, the user simply sets up the run with \texttt{basilic\_setup GW231123.yml}. This generates the folder structure of the analysis, stores data, PSD, and priors\footnote{Modification of default analysis priors can be done after running the setup, with the \texttt{basilic\_modify\_priors} command, specifying setup directory, model names, and prior paths.}, and sets up the Condor submission files, in the case of a single analysis, a \texttt{.sub} file, and in the case of an injection campaign, a \texttt{.dag} file. Submission to Condor is simply done by running the \texttt{basilic\_submit\_jobs} followed by the setup directory path. Finally, after Condor accepts, runs, and returns the populated result folders to the interactive node, the user can run \texttt{basilic\_postproc} to automatically create a more curated result version in the form of a \texttt{summary\_results.html}, which creates an all-in-one run report with plots of PSD used, per-model SNR distributions, parameter posteriors, Bayes factor statistic, and more. Alternatively, even without running the post-processing, the \texttt{result\_bilby} folder in the outdir contains the standard sampler plus \texttt{bilby} outputs, which might be sufficient depending on the user's needs for single-run analyses. For injection campaigns, that standardized output will become unwieldy, so post-processing becomes a vital step in interpreting the results. An example, sub-part of a standard \textsc{HTML} report produced by \Basilic~in injection campaign mode can be found in Fig.~\ref{fig:screenshot}.

\section{Morphological analysis}\label{app:pdeg}

As injection campaigns are such an integral part of \Basilic, we will address here explicitly how injections can be used to address model confusion and provide context for the interpretation of Bayes factors  and potential aid in model selection. In particular, we are interested in scenarios where the between-models Bayes factors are unconclusive ($\log_{10}\mathcal{BF}\sim1-3$), while the model versus noise Bayes factors are moderate or higher ($\log_{10}\mathcal{BF}\ge3-6$). 
This Appendix describes in detail the implementation of the two-stage diagnostic approach mentioned in Section~\ref{sec:discussion}. Stage I performs PPCs with the aim of quantifying the level of agreement between the proposed models potential predictions, and the actual observed data. For a motivation of PPC in a Bayesian context, see~\cite{gelman2013philosophy}. Our main motive for suggesting PPC in a burst model selection scenario is driven by the possibility of model misspecification. However, we acknowledge that abnormal Bayesian p-values are not only caused by poor hypothesis selection (relative to the observation), but also by likelihood assumptions (here Gaussianity and noise stationarity) and parameter prior selection. Stage II, invoked when PPC also does not distinguish the hypotheses through close Bayesian p-values, quantifies if the posterior-supported waveform families themselves are morphologically degenerate using a match-based existence criterion. If within these regions, the predicted waveforms of the considered models highly match, no observational qualities (through individual noise realizations) will permit model selection.

Stage I proceeds after the Bayesian model selection, by utilising the posterior samples \(\{\boldsymbol{\theta}_k^{(i)}\}_{i=1}^{N}\) from
\(p(\boldsymbol{\theta}_k\mid d^\ast,\mathcal{M}_k)\). The per-model Bayesian p-value is then estimated in practice by

\begin{equation}
      \widehat p^{\mathrm{ppc}}_k
  =
  \frac{1}{N}\sum_{i=1}^{N}
  \mathbf{1}\!\left\{
    T_k(d^{\mathrm{rep},(i)}) \ge T_k(d^\ast)
  \right\}
  \label{eq:pcc_pvalue}
\end{equation}

Computing \(T_k(d)\) given in Eq.~\ref{eq:ppc_Tk} requires exact evidence evaluations for each replicate, which
can be expensive; in practice, one may use coarse evidence settings for PPC, but the lower dimensionality of many burst models does permit such analyses.

Stage II is invoked if $\widehat p^{\mathrm{ppc}}_k$ does not clearly disfavor either hypothesis; This is for instance the case where similar Bayesian p-values are obtained because the two models $M_1$ and $M_2$ are morphologically degenerated:
the two waveform families overlap substantially in the region of parameter space
supported by the posteriors. This test is particularly important if, in addition to an inconclusive Bayes factor and PPC, the involved SNRs are high. We therefore quantify degeneracy directly at the waveform level, using posterior draws
and the GW match. For waveforms \(h_1(\boldsymbol{\theta}_1)\) and \(h_2(\boldsymbol{\theta}_2)\), the normalized overlap is
\begin{equation}
  O(\boldsymbol{\theta}_1,\boldsymbol{\theta}_2)
  :=
  \frac{\langle h_1(\boldsymbol{\theta}_1)\mid h_2(\boldsymbol{\theta}_2)\rangle}
  {\sqrt{\langle h_1(\boldsymbol{\theta}_1)\mid h_1(\boldsymbol{\theta}_1)\rangle\,
         \langle h_2(\boldsymbol{\theta}_2)\mid h_2(\boldsymbol{\theta}_2)\rangle}},
  \label{eq:overlap_def}
\end{equation}
and the (maximized) match
\begin{equation}
  \mathcal{M}(\boldsymbol{\theta}_1,\boldsymbol{\theta}_2)
  :=
  \max_{\Delta t,\Delta\phi}\,
  O\!\bigl(\boldsymbol{\theta}_1,\boldsymbol{\theta}_2\mid\Delta t,\Delta\phi\bigr),
  \label{eq:match_def}
\end{equation}
where \((\Delta t,\Delta\phi)\) represent extrinsic nuisance shifts (time and phase). The inner product is $\langle a\mid b\rangle
  :=
  4\,\Re\!\int_0^\infty \frac{\tilde a(f)\tilde b^\ast(f)}{S_n(f)}\,df$. 
  
Let \(m_0\in(0,1)\) be a chosen high match threshold (e.g.\ \(m_0=0.97\) or \(0.99\)).
For a fixed \(\boldsymbol{\theta}_1\), we define a binary degeneracy flag that depends on whether
there exists at least one \(\boldsymbol{\theta}_2\) in the posterior-supported region of
\(M_2\) producing a high match
\begin{equation}
  \mathrm{Ind}_{1\to2}(\boldsymbol{\theta}_1\mid m_0)
  :=
  \mathbf{1}\!\left\{
    \sup_{\boldsymbol{\theta}_2}
    \mathcal{M}(\boldsymbol{\theta}_1,\boldsymbol{\theta}_2)
    \ \ge\ m_0
  \right\}.
  \label{eq:deg_exist_sup}
\end{equation}

We then integrate this flag over the posterior of \(\boldsymbol{\theta}_1\)
\begin{align}
  P_{1\to2}^{\mathrm{match}}(m_0)
  :&=
  \Pr_{\boldsymbol{\theta}_1}
  \!\left(
    \mathrm{Ind}_{1\to2}(\boldsymbol{\theta}_1\mid m_0)=1
  \right)\\
  &=
  \int \mathrm{Ind}_{1\to2}(\boldsymbol{\theta}_1\mid m_0)\,
  p(\boldsymbol{\theta}_1\mid d^\ast,\mathcal{M}_1)\,d\boldsymbol{\theta}_1.
  \label{eq:P12_match}
\end{align}
This quantity has the desired limiting behavior:
if \(M_1=M_2\) and their posteriors coincide, then
\(P_{1\to2}^{\mathrm{match}}(m_0)=1\) for any \(m_0<1\); if the waveform manifolds are
``posterior-orthogonal'' (no high matches exist), then \(P_{1\to2}^{\mathrm{match}}(m_0)=0\).

Because degeneracy is inherently a claim about both families, and to account for potentially poorly constrained posteriors, we also compute the reverse
direction \(P_{2\to1}^{\mathrm{match}}(m_0)\) and report a symmetric quantity
\begin{equation}
  P_{\mathrm{deg}}^{\mathrm{match}}(m_0)
  :=
  \min\!\left\{
    P_{1\to2}^{\mathrm{match}}(m_0),
    P_{2\to1}^{\mathrm{match}}(m_0)
  \right\},
  \label{eq:Pdeg_match_sym}
\end{equation}
The minimum is a conservative choice that prevents one
broad posterior from trivially implying degeneracy if the reverse direction does not
support it. If the interest in one model's parameters and posterior is higher, the reverse can be ignored as the result from $P_{1\to2}^{\mathrm{match}}(m_0)$ would still imply that there is degeneracy driving the inconclusive Bayes factors observed.

In practice, as with PPC, we have \(\{\boldsymbol{\theta}_1^{(i)}\}_{i=1}^{N_1}\), the posterior samples from
\(p(\boldsymbol{\theta}_1\mid d^\ast,\mathcal{M}_1)\) and \(\{\boldsymbol{\theta}_2^{(j)}\}_{j=1}^{N_2}\) samples from
\(p(\boldsymbol{\theta}_2\mid d^\ast,\mathcal{M}_2)\). The empirical search statistic
\begin{equation}
  \widehat{\mathcal{M}}_{\max}(\boldsymbol{\theta}_1^{(i)})
  :=
  \max_{1\le j\le N_2}\,
  \mathcal{M}\!\left(\boldsymbol{\theta}_1^{(i)},\boldsymbol{\theta}_2^{(j)}\right),
  \label{eq:match_max_empirical}
\end{equation}
and the estimated flag
\begin{equation}
  \widehat{\mathrm{Ind}}_{1\to2}(\boldsymbol{\theta}_1^{(i)}\mid m_0)
  :=
  \mathbf{1}\!\left\{
    \widehat{\mathcal{M}}_{\max}(\boldsymbol{\theta}_1^{(i)}) \ge m_0
  \right\}.
  \label{eq:deg_flag_empirical}
\end{equation}
Then
\begin{equation}
  \widehat P_{1\to2}^{\mathrm{match}}(m_0)
  =
  \frac{1}{N_1}\sum_{i=1}^{N_1}
  \widehat{\mathrm{Ind}}_{1\to2}(\boldsymbol{\theta}_1^{(i)}\mid m_0),
  \label{eq:P12_match_hat}
\end{equation}
and similarly for \(\widehat P_{2\to1}^{\mathrm{match}}(m_0)\) and
\(\widehat P_{\mathrm{deg}}^{\mathrm{match}}(m_0)\). The dominant cost is the number of
match evaluations, here $\mathcal{O}(N_1N_2)$, which is why we do not actually maximize over $\boldsymbol{\theta}_2$ for each $\boldsymbol{\theta}_1$.

Both the PPC and waveform match check are implemented in the postprocessing module of \Basilic. These are optional checks that can be performed when the initial analysis' nested sampling return inconclusive Bayes factor and further model checking is warranted. This scheme has been proposed in an entirely posterior-dependent way, as in practice, the interest in models might be purely data-driven. Unlike BBH and their waveform approximants, burst signals are not only currently unobserved, but some also are attributed to astrophysical phenomena with no a posteriori support (e.g. cosmic strings). Therefore this scheme is aimed at future potential burst-like signals and current sub-threshold GW events where model selection alone might not be entirely satisfactory scientifically.


\bibliography{apssamp}

\end{document}